\documentclass[aip,amsmath,amssymb,preprint]{revtex4-1}

\usepackage{bm}
\usepackage{natbib}
\usepackage{graphicx}
\usepackage{epstopdf}
\usepackage{hyperref}
\usepackage{color}
\usepackage{soul}
\usepackage{cleveref}
\usepackage{caption}
\usepackage{amssymb,amsfonts,amsmath}

\DeclareFontFamily{U}{euc}{}
\DeclareFontShape{U}{euc}{m}{n}{<-6>eurm5<6-8>eurm7<8->eurm10}{}
\DeclareSymbolFont{AMSc}{U}{euc}{m}{n} 
\DeclareMathSymbol{\umu}{\mathord}{AMSc}{"16} 
 
\begin{document}

\newcommand{\fdrive}{f_\text{drive}}
\newcommand{\rbead}{r_\text{B}}
\newcommand{\rdimer}{R_\text{p}}
\newcommand{\wi}{W \! i}
\newcommand{\degreesC}{\,^{\circ}{\rm C}}
\newcommand{\degrees}{\,^{\circ}}

\title{{\color{black}Flagellar} Kinematics and Swimming of Algal Cells in Viscoelastic Fluids}
\author{B. Qin} 
\affiliation{
Department of Mechanical Engineering \& Applied Mechanics, University of Pennsylvania,
Philadelphia, PA 19104
}
\author{A. Gopinath}
\affiliation{
Department of Physics \& Astronomy, Haverford College, Haverford, PA
19041
}
\affiliation{
Department of Mechanical Engineering \& Applied Mechanics, University of Pennsylvania,
Philadelphia, PA 19104
}
\author{J. Yang}
\affiliation{
Department of Physics \& Astronomy, Haverford College, Haverford, PA
19041
}
\affiliation{
Department of Mechanical Engineering \& Applied Mechanics, University of Pennsylvania,
Philadelphia, PA 19104
}
\author{J. P. Gollub}
\affiliation{
Department of Physics \& Astronomy, Haverford College, Haverford, PA
19041
}
\author{P. E. Arratia}
\email[]{Author to whom correspondence should be addressed. Electronic mail: parratia@seas.upenn.edu.}
\affiliation{
Department of Mechanical Engineering \& Applied Mechanics, University of Pennsylvania,
Philadelphia, PA 19104
}

\begin{abstract}
{The motility of microorganisms is influenced greatly by their hydrodynamic interactions with the fluidic environment they inhabit. We show by direct experimental observation 
of the bi-flagellated alga \emph{Chlamydomonas reinhardtii} that fluid elasticity and viscosity strongly influence the beating pattern - the gait - and thereby control the propulsion speed. The beating frequency and the wave speed characterizing the cyclical bending are both enhanced by fluid elasticity. Despite these enhancements, the net swimming speed of the alga is hindered for fluids that are sufficiently elastic. The origin of this complex response lies in the interplay between the elasticity-induced changes in the spatial and temporal aspects of the flagellar cycle and the buildup and subsequent relaxation of elastic stresses during the power and recovery strokes.} 
\end{abstract}

\maketitle

The motility of microorganisms and organelles through microstructured fluids plays an important role in varied biological processes such as fertilization and genetic transport\cite{Machin:1958,Katz1978, Suarez:2001,fauci_2006}, development of disease \cite{Celli2009} and bio-degradation in soil \cite{Alexander1991}. Disruption of normal motility can occur due to unexpected changes in the nature of the fluids. For instance, the graceful beating of filamentous cilia pumping mucus in the respiratory tract\cite{Lillehoj2002, fauci_2006, Lai2009} and the flagella driven swimming of spermatozoa through cervical mucus \cite{Suarez:1992, Guzick2001} are both affected by the properties of the mucus such as water content and viscoelasticity. At larger scales, the undulatory motion of {\em C. elegans} in wet soil \cite{juarez_2010} or through polymer networks\cite{Gagnon2013} is influenced by the rheology of the environment sensed by the worm as it moves.  In addition, there is growing interest in designing artificial swimmers driven by external fields\cite{dreyfus_2005,keim_2012} and developing objects capable of swimming remains a subject of  intensive exploration.

Many organisms move in the realm of low Reynolds number  Re $\equiv$ $\rho \ell U/\mu \ll 1$ because of either small length scales $\ell$, low swimming speeds $U$ or both.  In a Newtonian fluid with constant density $\rho$ and viscosity $\mu$, this implies that inertial effects are negligible and that the stresses felt by the swimmer are purely viscous and linear in the viscosity. Since the Stokes equations govern the fluid response to the moving body, the fluid kinematics is reversible. To therefore swim, organisms must execute non-reversible, asymmetric strokes in order to break free of the constraints imposed by the so-called ``scallop theorem."~\cite{purcell_1977} In many instances, however the ambient fluid environment is far from Newtonian due to the presence of macromolecules such as biopolymers and proteins, which impart complex rheological characteristics such as shear rate dependent viscosity and viscoelasticity. In a viscoelastic fluid, stresses are both viscous and elastic, and therefore time dependent - consequently, kinematic reversibility can break down. This effect is especially important for small organisms since the relaxation time of elastic stresses in the fluid $\lambda$ may then become comparable to the viscous diffusion time of vorticity  $\rho \ell^2/\mu$. For microorganisms with cyclical swimming strokes, the elastic stresses may then persist between periods and eventually dominate over viscous effects.

The consequences of fluid elasticity on the details of swimming while clearly important, are not well-understood, and {\color{black}have} therefore received growing attention.~\cite{fauci_2006, Fu:2009, Lauga:2007, teran_2010, Fu:2007, Shen:2011, pak_2012, keim_2012, liu_2011, becca2014} Most recent work has been theoretical in nature relying on detailed simulations~\cite{teran_2010, Lauga2012, becca2014} or asymptotic solutions to idealized models.~\cite{Fu:2009, Lauga:2007, Fu:2007} Systematic experiments that may shed light are scarce and mainly focussed on swimming in Newtonian fluids \cite{berg_1979, drescher2010, guasto_2010}. Taken together, current studies paint a complicated and sometimes contradictory picture. For instance, theories on the small amplitude swimming of infinitely long wave-like sheets suggest that elasticity can reduce swimming speed \cite{Fu:2009,Lauga:2007} and these predictions are consistent with experimental observations of undulatory swimming in \textit{C. elegans}.\cite{Shen:2011}. Similar trends were found recently\cite{Lauga2012} on studies of the motility of both idealized ``pullers" (such as \emph{C. reinhardtii}) and ``pushers" (such as \emph{E. coli}). On the other hand, simulations of finite-sized moving \cite{teran_2010} filaments or large amplitude undulations \cite{becca2014} suggest that fluid elasticity can increase the propulsion speed  - consistent this time with experiments on propulsion due to rotating rigid mechanical helices.\cite{liu_2011} The emerging viewpoint is that fluid microstructure and swimming kinematics {\em together} impact motility in a non-linear manner\cite{Gagnon2013, becca2014}.

In this manuscript, we experimentally investigate the effects of fluid elasticity on the swimming behavior of the bi-flagella green alga, \emph{Chlamydomonas reinhardtii}. With an ellipsoidal cell body (Fig. 1a) that is roughly 10 $\mu$m in size and  two anterior flagella each of length $\ell \sim$ 10 $\mu$m, the alga {\em C. reinhardtii} is a model system in biology\cite{harris_1999} and has been widely used in studies on motility. The two flagella possess the same conserved ``9+2'' microtubule arrangement seen in other eukaryotic axonemes \cite{harris_1999}  and as a pair, execute cyclical breast-stroke like patterns with asymmetric power and recovery strokes at frequencies varying from $\omega \sim $ 30-60 Hz to generate propulsion.\cite{guasto_2010,harris_1999} This swimming gait generates far-field flows corresponding to an idealized ``puller." \cite{Lauga2012, drescher2010} By systematically modifying the elasticity of the fluid, we studied  the variation of the flagellar beat pattern, beat frequency $\omega$ and centroid velocity. We find that fluid elasticity can modify the  beating pattern (i.e. shape) and enhance the alga's beating frequency and wave speed. Despite this enhancement, the alga's swimming speed is overall hindered (as much as 50\%) by fluid elasticity due to the elastic stresses  in the fluid. 

Two types of fluids are used -- Newtonian and viscoelastic. Newtonian fluids are prepared by dissolving relatively small quantities of Ficoll (Sigma-Aldrich) in M1 buffer solution. The Ficoll concentration in M1 buffer is varied from 5\% to 20\% by weight in order to produce fluids with a range of viscosities (1 cP to 10 cP). Viscoelastic fluids are prepared by adding small amounts of the flexible, high molecular weight polymer polyacrylamide (PAA, MW=$18 \times 10^6$, Polysciences) to water. The polymer concentration in solution ranges from 5 to 80 ppm resulting in fluid relaxation times $\lambda$ that range from 6~ms to 0.12~s, respectively (SI-$\S$I, Figure 3a, Table 1). These polymeric solutions are considered dilute since the overlap concentration for PAA is approximately 350 ppm. All fluids are characterized using a cone-and-plate rheometer (RF3, TA Instruments); for more information, please see SI-$\S$I. {We use an effective viscosity $\mu = \left[ \eta(\dot{\gamma}_{\mathrm{body}})+ \eta( \dot{\gamma}_{\mathrm{flag}})\right]/2 $ to study viscous effects on swimming in the polymeric fluids (Methods and Materials), where $\eta(\dot{\gamma})$ is the shear rate dependent viscosity. This enables us to compare swimming in these fluids to swimming in a Newtonian fluid with the same effective viscosity.}

Dilute {\em C. reinhardtii} suspensions are made by suspending motile algae in either Newtonian (Ficoll) or viscoelastic (PAA) fluids. A small volume of this suspension is then stretched to form a thin film (thickness $\approx$ 20 $\mu$m) using a wire-frame device (Methods and Materials, SI-$\S$I,A), and cell motion in the thin film is imaged using an optical microscope and a high speed camera. The cell centroid and flagella of cells that exhibited regular synchronous swimming strokes (for other observed gaits see SI-$\S$II and SI-Movies) are identified and tracked. Figure 1(a) shows snapshots of \emph{C. reinhardtii} swimming at Re $\sim10^{-3}$ in a water-like M1 buffer solution. The shapes of the flagella (red curves - Fig. 1a) as well as the trajectory of the cell centroid are tracked simultaneously with instantaneous swimming speeds calculated by differentiating the centroid position and the sign determined from the cell orientation. The speed of the swimmer measured in the thin liquid film set-up are consistent with that in the bulk fluid reported by other researchers.\cite{guasto_2010}

The beating pattern of \textit{C. reinhardtii} over one cycle in Newtonian (N) and polymeric solutions (VE) of similar viscosity $\mu$~$\approx$~6 mPa$\cdot$s are presented in Fig. 1(b) (c.f. SI-$\S$II for analogous shapes for $\mu$~$\approx$~2.6 mPa$\cdot$s). The difference in shapes is striking and illustrates the effects of fluid elasticity on swimming. In the Newtonian case (i), the flagellum seems more mobile and significant changes in curvatures are attained over the whole cycle. In the viscoelastic case (ii), lateral displacements of almost a third of the flagellum close to the cell body (green) appear to be severely restricted (less mobile) or bundled together with most of the bending occurring over the remainder of the length. Furthermore, we observe localized bending at the distal tip in the initial stages of the power stroke. The differences in the shapes can be quantified by plotting the spatio-temporal evolution of the scaled flagellum curvature $\kappa(s, t) = \tilde \kappa(s, t) \cdot\ell$ over many cycles (Fig. 1c). These kymographs show that regions of high curvature are found to distribute diagonally and periodically, characteristic of propagating bending waves.  In the elastic fluid, (Fig. 1c, right panel) flagella attain larger curvatures  (dark blue regions) and an increase in the frequency of bending waves (diagonally oriented lines - direction shown by arrow). We also observed that for very low viscosities ($1 < \mu < 2.6$ mPa$\cdot$s), the distal tip gets closer to the cell body than for higher viscosity fluids (SI, Fig. 7). {We can quantify this difference in curvature by computing the normalized curvature averaged over time $t$ ($\approx 6$ cycles) and dimensionless arc length $s/\ell$, here denoted by $\langle \kappa\rangle$. We find that at $\mu \approx 6$ cP, the value of $\langle \kappa\rangle$ is -1.44 and -2.72 for the Newtonian and viscoelastic fluids, respectively.}

These changes in the spatio-temporal dynamics of the flagella translate to variations in beat frequency and swimming speed. We begin by investigating the effects of shear viscosity on the beat frequency (Fig. 2a) and on the cycle averaged net swimming speed (Fig. 2b).  In Newtonian fluids, for viscosities $\mu \sim 2$ $\mathrm{mPa\cdot}$s and lower, the frequency is roughly around 56 Hz. Increasing the viscosity further ($\mu > 2$ $\mathrm{mPa\cdot}$s) results in a monotonically decreasing frequency. Intriguingly, we find that the decay is well captured by $\omega \sim 1/\sqrt{\mu}$ consistent with models suggesting that emergent frequencies are selected based on a balance of internal active processes, the elastic properties of the flagellum and external viscosity\cite{Camalet2000} (see also SI-$\S$III).  For polymeric fluids with low viscosities (low PAA concentration), the frequencies are consistent with the Newtonian values. At higher concentrations of polymer, however, significant deviation from the Newtonian trend is observed, even when the fluid has comparable viscosity. Surprisingly, the beating frequency increases with increasing fluid viscosity and then seems to saturate. {Note  from the data that the variation in beating frequencies at each characteristic fluid/relaxation time is small as seen from the relative size of the error bars compared to the mean value. }

The observed increase in beating frequency, however, does not translate into an increase in overall swimming speed. Figure 2(b) shows that the average net swimming speed $U$ of the \emph{C. reinhardtii} cell body decreases as the fluid viscosity increases for both Newtonian and polymeric solutions. For low viscosity values ($\mu < $~2~$\mathrm{mPa\cdot}$s), the swimming speeds are very similar for the algae cells in Newtonian and polymeric liquids. As polymer concentration increases, however, we find that fluid elasticity consistently hinders self-propulsion compared to Newtonian fluid at comparable viscosity. For both Newtonian and polymeric fluids, we find the data consistent with the relationship $U\sim \mu^{-1}$, which suggests that the algae are operating at nearly constant thrust. Such a relationship has also been observed for free swimming \emph{C. reinhardtii} in low viscosity Newtonian fluids ($1<\mu< 2$~$\mathrm{mPa\cdot}$s)~\cite{Peyla2010}.  

The data shown in Fig. 2 emphasize the increasing importance of elasticity as the polymer concentration in the fluid increases. Viscoelastic effects can be quantified by introducing the Deborah number, defined here as De $\equiv \omega \lambda$ where $\omega$ is the mean frequency of a representative sample of cells and $\lambda$ is the fluid relaxation time; we note that De $=0$ for Newtonian fluids. The normalized beating frequency and algal swimming speed as a function of De are shown in Fig. 3(a) and 3(b), respectively. For ease of presentation, we normalize the measured frequency and net swimming speed with the Newtonian value at comparable viscosity. Figure 3(a) shows the monotonic increase in the frequency as the relaxation time of the fluid increases from around 20 to 120 ms. The transition from a Newtonian-like response (De $\lesssim 1$) to a clear viscoelastic regime occurs at around De $\sim 2.5$, where the frequency is on the order of the fluid relaxation time scale, suggesting that elastic fluid stresses are modifying kinematics. The ratio of swimming speeds plotted in Fig. 3(b) is consistently less than unity demonstrating that fluid elasticity hinders net locomotion. The decrease is quite substantial even for relatively low values of De. For example, fluid elasticity hinders the cell swimming speed, relative to Newtonian fluids, by as much as 50\% for De $\approx 2$. We also observe that the ratio plateaus to approximately 0.4 for De $> 2$. This asymptotic behavior has been previously observed in theoretical studies \cite{Lauga:2007, Fu:2009} and also in experiments with worms.\cite{Shen:2011} The reduction in motility is also consistent with recent simulations~\cite{Lauga2012} of steady flow of weakly elastic fluid around idealized pullers. This plateau may indicate an upper bound on the generated elastic stress around the organism. 

The net frequency and swimming speed while central to the alga's overall motility, do not distinguish between the highly asymmetric power and recovery strokes. Therefore, we next calculate the mean speed during the power stroke $U^{+}$ and the mean speed during the recovery stroke $U^{-}$ for Newtonian and viscoelastic fluids. Figures 4(a) and 4(b) summarize our observations. For Newtonian fluids, both the power $U^{+}$ and recovery $U^{-}$ stroke speeds decrease as $\mu$ increases, following the trend observed earlier for the Newtonian net swimming speed in Fig. 3(b). For viscoelastic fluids, however, we observe a sharp difference between power and recovery strokes. While the dependence of $U^{+}$ is similar to the Newtonian case, the recovery speed $U^{-}$ in viscoelastic fluids remains relatively unchanged, and in fact modestly increases with viscosity. This raises the possibility that fluid elasticity affects power and recovery strokes very differently -  a view supported by plots of the normalized power and recovery stroke speeds shown in Fig 4(c) and 4(d). We observe a minimum in the values of $U_{\mathrm{VE}}^{+}/U_{\mathrm{N}}^{+}$ and $U_{\mathrm{VE}}^{-}/U_{\mathrm{N}}^{-}$  at very low values of De. For larger values of De (De $>2$), the speed ratio during power stroke, $U_{\mathrm{VE}}^+/U_{\mathrm{N}}^+$ starts from values less than unity, but then increases and in fact exceeds unity at De $=6.4$. On the other hand, the speed ratio during the recovery stroke, $U_{\mathrm{VE}}^-/U_{\mathrm{N}}^-$ is consistently greater than unity and increases with De.  

Seeking signatures of this complex response in the coupling between cell motion and flagellar strokes, we revisit the beat patterns illustrated in Fig. 1(c).  As anticipated, we find a significant difference in the shape of the flagella (Fig. 5a) at the onset of the power stroke.  For the viscoelastic case, the bending at the distal end (green arrow) is much more pronounced, consistent with the tip flexion observed in the contours in the inset of Fig. 5(c). Also, there are pronounced curved regions at both the proximal and distal end, while the profile for the Newtonian case is relatively flatter. We recall that as De increases, flagellar displacements over a cycle at the proximal end become increasingly confined, yielding a ``bundled'' shape that is evident for De $=2.4$ (SI-$\S$II) and more prominent for De = $6.5$ (inset in Figs. 5b and 5c). We quantify these observations in Figs. 5(b) and 5(c) by plotting mean curvatures, calculated separately for the power cycle (blue) and recovery cycle (red), and comparing them to the cycle averaged curvature (black).  For the viscoelastic case (Fig. 5c), the curvature is relatively fixed at $s\approx 0$ throughout the cycle, while for the Newtonian case (Fig. 5b), the curvatures vary significantly and in fact flip signs over a beat cycle. 
These shapes are suggestive of localized polymer stress regions at both the distal and proximal end. 

Such regions of concentrated stress have been predicted\cite{becca2014} near both ends of a free undulatory burrowing swimmer in polymeric fluids. In our case, these regions may develop due to the curvature of fluid streamlines induced by vortices that strengthen and shift towards the anterior of the cell body during the power stroke\cite{guasto_2010}. Furthermore, the boundary effects may amplify these stresses near the proximal end $s\approx 0$. We hypothesize that the proximal elastic stress concentration has opposing effects for the power and the recovery stroke. During the power phase, signed speed data show that the cell body is pulled forward by the flagella. This forward motion is resisted by the proximally located elastic stresses in the fluid, hence the motion will be impeded compared to the Newtonian case - i.e,  $U_{\mathrm{VE}}^+/U_{\mathrm{N}}^+ < 1$ as seen for $2<$ De $<6$ in Figure 4(c). Added to this, fluid stretching at the hyperbolic flow points\cite{guasto_2010}  in the anterior of the cell body results in enhanced extensional viscosity that hinders this forward motion. On the other hand, when the cell body recoils during the recovery phase, the relaxing polymer stress amplifies the backward motion, augmenting the speed  - and thus $U_{\mathrm{VE}}^+/U_{\mathrm{N}}^+ > 1$. The increase and build up of elastic stresses at the distal end may also separately enhance the power and recovery strokes by preventing the swimmer from slipping backwards as observed in simulations\cite{teran_2010}. Consistent with this picture, we observe in Fig 5(b) that as De increases, the backward speed during the recovery stroke increases. For the forward stroke, however, we {\rm do not} see a monotonic drop in $U_{\mathrm{VE}}^+/U_{\mathrm{N}}^+$ with Deborah number, as one would expect by considering elastic stress alone. Rather, the ratio increases for De$\ge 2$ and eventually attains values greater than unity. 

A closer examination of the time scales of the  beat  cycle provides clues to a possible explanation. We note that the increase in $\omega_{\mathrm{VE}}/\omega_{\mathrm{N}}$ at De $\approx 2$ coincides with the increase in the power stroke speed.  For beating at constant  stroke pattern - thus constant amplitude - the speed may be expected to increase with the frequency in a manner similar that for idealized swimming sheets\cite{Fu:2007,Lauga:2009}.  Here, however, the increase in frequency is accompanied by a change in both the beating amplitude and flagellar curvatures, c.f. Fig. 5(a-c). Together, these have striking effects on the temporal composition of the beat cycle and resultant cell body displacements.  In Figures 5(d), we show instantaneous speeds in Newtonian fluids with two different viscosities plotted as a function of $t/T$, where $T$ is the respective beating period. We observe the expected self-similarity where the instantaneous speed profiles collapse up to an amplitude scaling. Peak speeds of the recovery stroke occur at integer values of $t/T$; zero speed points coincide indicating equal proportion of power phase and recovery phase for beat patterns. Comparison of the Newtonian case with its viscoelastic analogue reveals a richer picture, c.f. Figure 5(e). The viscoelastic case has slightly higher amplitudes for both power and recovery stroke. This is possibly due to polymer stretching ``overshoot'' that occurs at the switch between the power and recovery stroke, as observed in the transient extensional flows in dilute polymers\cite{feng2000}. Furthermore, the zero speed points coincide only at the start of the recovery stroke and the consistent delay in the start of the power stroke shows that the alga is experiencing a more extended recovery and a shortened power phase. 

Overall, our experiments show that during the recovery stroke, the polymer stress in the fluid, the increase in beating frequency and the increase in the proportion of recovery stroke work in concert to enhance the speed during recovery stroke.  In contrast, during the power stroke, the three effects compete with each other.  The net result ultimately yields a reduction in speed of the alga. 

{In this work, we focus on the algae {\em C. reinhardtii} but the feedback we observe between fluid elasticity and swimming strokes is a more general principle. Many live organisms and organelles live and move in fluids that possess both fluid-like and solid-like behavior (i.e. viscoelasticity), including mammalian spermatozoa progressing in human mucus (of different viscosities)~\cite{Katz1978, Guzick2001}, nematodes burrowing in wet soil~\cite{juarez_2010}, and cilia beating in the respiratory tract \cite{Lillehoj2002}. It is now recognized that there is no straightforward relationship between fluid elasticity and motility.\cite{fauci_2006,teran_2010,becca2014} Rather, the motility behavior of organisms seems to depend on the swimmer's kinematics and fluid properties as well as the material properties of the the swimmer itself. In fact, recent numerical results show that softer organisms respond differently to fluid elasticity than stiff organisms~\cite{becca2014}. Also, our results show that fluid elasticity influences kinematics and swimming speeds in a manner different from just shear thinning viscosity, which for instance cannot account for the non-monotonic dependence of beating frequency on polymer concentration or viscosity (Fig. 2a) as well as the dramatic increase ($\approx$ 50\%) in frequency for the polymeric solution compared to the Newtonian case at similar viscosity ($\mu=5.7$cP). Indeed, recent theoretical analysis \cite{lauga2013} and experimental results with {\it C. elegans} \cite{gagnon2014} show little to no effect of shear thinning viscosity on swimming kinematics including propulsion speed.} Finally, it is not known whether swimmers can use mechanosensation to actively control their swimming modes and attempt to overcome retarding fluid effects. Incorporating and understanding the interplay between fluid stresses, the material properties of the active filament, and the forces/moments produced by such filaments is a natural next step to distinguish active and passive effects.

\section*{Methods and Materials}

\noindent{\bf Thin film and imaging}
\\\\
A small amount of motile alga was added to the solutions thus prepared and the suspension was further stabilized  by adding the surfactant Tween-20. A small volume of this suspension was stretched into a thin film of thickness $\approx$ 20 $\mu$m using an adjustable wire-frame device.~\cite{guasto_2010} The film thickness is about twice the alga body diameter so the cells are unable to rotate about their swimming axis and only swimming in the mid plane of the film was recorded.  The film interfaces are nearly stress-free which minimizes velocity gradients transverse to the film.~\cite{guasto_2010} 
Cell motion in the thin film is imaged using a microscope (dark field, 20x or 60x magnification) and a high-speed camera (600 fps). We expect no interfacial effects as the flagella beat many (flagellum) diameters away from the interface. {\color{black}Further, waveform data are collected only when cell body remains in focus and the optical signal of the two flagella do not overlap or interfere. Hence the cell is swimming in plane and is not rotating around its orientation axes. Due to limitation in imaging frame rate and the greater noise-to-signal level at the last 5-10\% of the flagella in the distal end, we exclude this region from our swimming waveform analysis}. Typical beating frequencies $\omega$ were on the order of $30$-$60$ Hz and the Deborah number $0 \leq $ De $< 7$ with Re $\sim 10^{-3}$. Under these conditions, the dimensionless Elasticity number El $\equiv$ De/Re $\gg 1$,  so that elastic effects are significant and are indeed being probed. 
\\\\
\noindent{\bf Analysis of swimming trajectories}
\\\\
Standard particle tracking techniques are used to measure the {\em C. reinhardtii} centroid position. Raw images of the algae swimming first undergo image processing such as contrast enhancement for stronger signal on the body and band pass filtering for the removal of the high-frequency noise in the image. The centroid and the orientation of the cell body  are then obtained using a non-linear least square method akin to template matching. The magnitude of the instantaneous swimming speed (or signed speed) is the ratio of the magnitude of the displacement vector over the time step. The sign of the instantaneous speed, i.e. from forward to backward or vice versa, is determined by inspecting the angle between the velocity and the cell body orientation.  The net speed is the averaged speed averaged over multiple beating cycles (SI-$\S$II,A).
For each rheologically distinct fluid, we average over 10-20 sample individuals that move with gaits corresponding to regular beating wherein both flagella move symmetrically (see SI-\S II,A and also SI-Movies). We also observed other gaits that once identified using energy spectra and frequency peaks were discarded before analysis. 
\\\\
{In order to compare swimming in polymeric fluids with that in Newtonian fluids, we define an effective viscosity $\mu =\frac{1}{2}[\eta(\dot{\gamma}_{\mathrm{body}})+\eta(\dot{\gamma}_{\mathrm{flag}})]$. Here $\eta$ is the shear rate dependent viscosity, $\dot \gamma_{\mathrm{body}} \equiv |U|/D$ where $|U|$ is the average of the power and recovery stroke speeds and $D$ is the cell diameter, and $\dot{\gamma}_{\mathrm{flag}} \equiv \omega$, the beating frequency. This effective viscosity is the average between the minimum shear rate experienced by the body and the maximum shear rate experienced by the flagellum.
}
\\\\
\noindent{\bf Tracking flagellar positions}
\\\\
The movement of the flagella during the beating cycle are tracked first by manual identification after which pixel positions and intensities corresponding to the material points are extracted. We then discretized the shapes 
and used an iterative process applying standard algorithms to obtain smooth and differentiable contours with low relative errors ($< 1\%$). The shapes relative to the center of the cell body obtained from multiple sequential periods are then mapped to points along a cyclical trajectory of constant period (SI-$\S$II,B) - the final contours are then inspected for fidelity to the actual discrete data.


\section*{Acknowledgements}
\noindent
 We thank H. Hu, J. Guasto, G. Huber, X. Shen, N. Keim, D. Gagnon and A. Koser for insightful discussions, and  K. Johnson for providing wild-type \emph{Chlamydomonas}. This work was supported by the National Science Foundation - DMR-1104705.

\section*{Author contributions}
\noindent
B.Q., J. Y.  J. P. G. and P. E. A. designed experiments; J. Y. and B.Q. performed experiments; B.Q. , A. G.  and J. Y. analyzed experimental data; A.G, B. Q. and P. E. A. developed theories to interpret analysis; A.G, B.Q. and P. E. A. wrote the manuscript; all authors discussed and interpreted results; J. P. G and P. E. A. conceived and supervised the project.

\section*{Additional Information}
\noindent
Supplementary information is available in the online version of the paper. Correspondence and requests for materials should be addressed to P. E. A.

\section*{Competing financial interests}
\noindent
The authors declare no competing financial interests.
%
\section*{Figures}
\begin{figure*}[t]
\includegraphics[width=15.0cm]{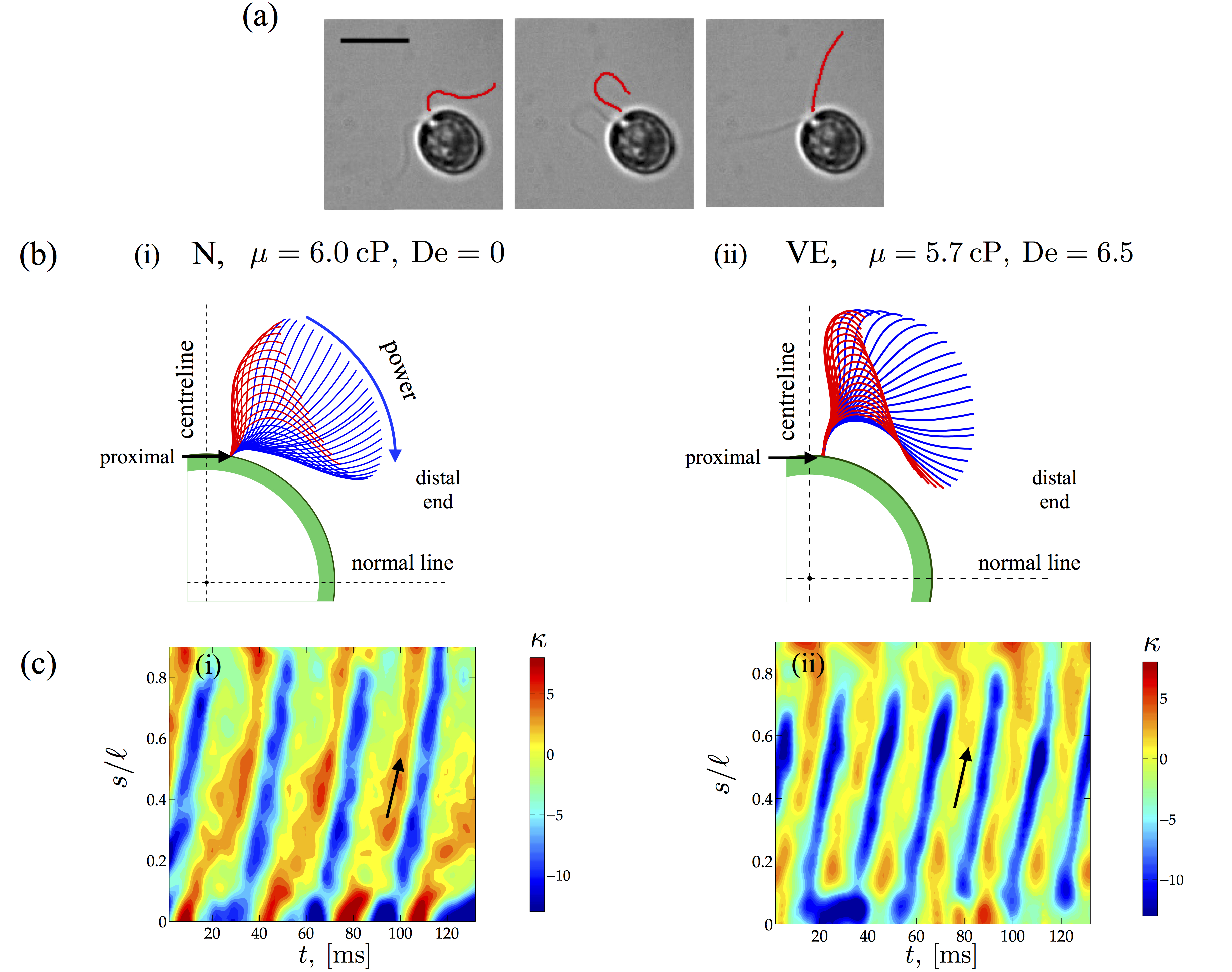}
\caption{ (Color online) {\bf Swimming kinematics.} (a) Sample snapshots of \emph{C. reinhadtii} swimming in buffer solution ($\mu=1 \:\mathrm{mPa}\cdot{\mathrm{s}}$); scale bar is 10 $\mu$m. We track the centroid of the cell body as well as trace the flagella over many oscillations. Red curves represent instantaneous flagellar shapes traced at various points in its beating cycle. Panels (b) Typical  contours for one complete cycle illustrating the shapes during power (blue) and recovery strokes (red). These shapes correspond to (i)  Newtonian fluid (N, De $=0$, $\mu = 6.0$ mPa$\cdot$s) and (ii) viscoelastic fluid (VE, De $=6.5$, $\mu = 5.7$ mPa$\cdot$s). (c) Kymographs of the spatio-temporal normalized curvature, $\kappa(s,t)$ along a flagellum for (i) Newtonian fluid - $\mu = 6.0$ mPa$\cdot$s and (ii) viscoelastic fluid  - $\mu = 5.6$ mPa$\cdot$s, De $= 6.5$. We see that increasing the elasticity results in larger curvature magnitudes (darker blue regions), significant differences in curvatures at the proximal end $s\approx 0$, and an increase in the frequency of bending waves (diagonally oriented lines - direction shown by arrow).}
\label{fig:1}
\end{figure*}
\begin{figure*}[t]
\includegraphics[width=15.0cm]{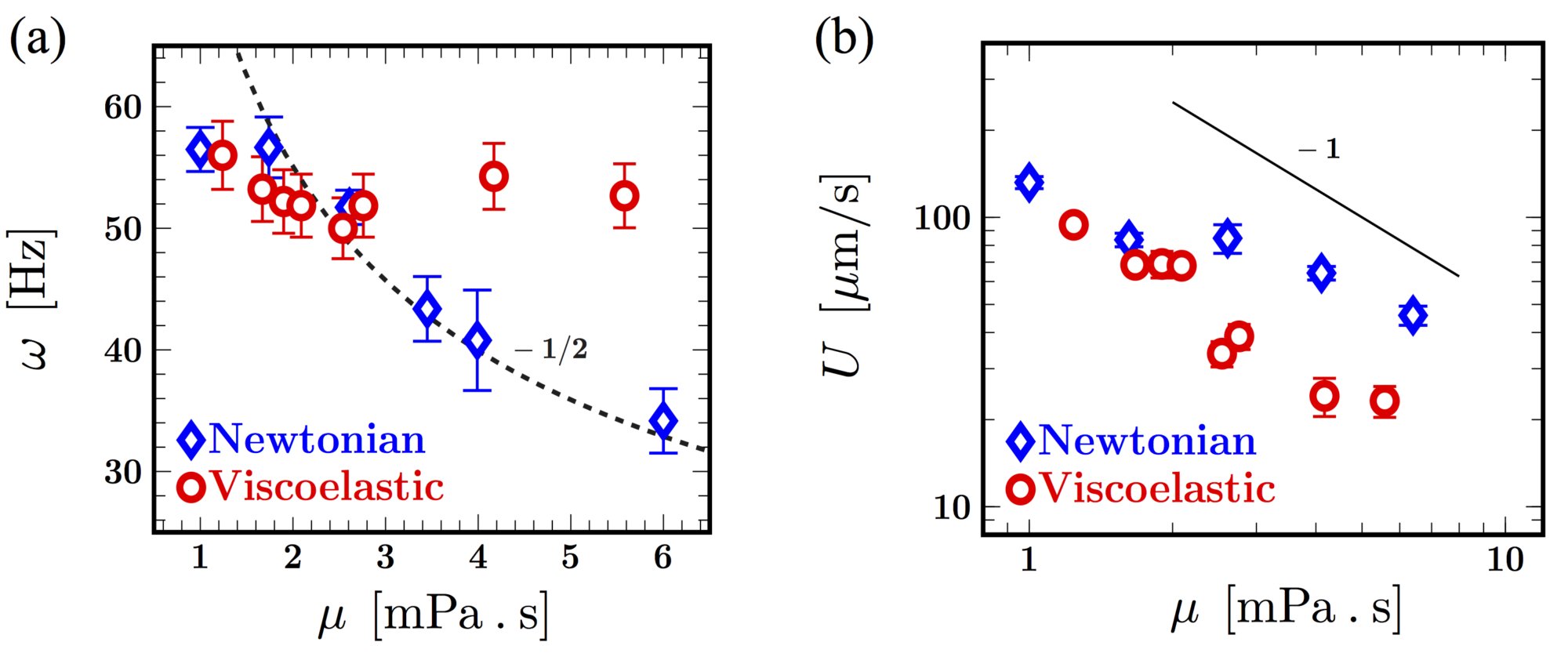}
\caption{ (Color online)  {\bf Effect of viscosity.} Effect of increasing viscosity on the beat frequency $\omega$ and swimming speed $U$ in Newtonian (blue, diamonds) and viscoelastic fluids (red, circles). (a) The beating frequency $\omega$ monotonically decays with viscosity for Newtonian fluids. For viscoelastic fluids, we observe two clear regimes. For low viscosities, the trend follows that for a Newtonian fluid. At higher viscosities - beyond $\mu \sim 2.6$ mPa.s, we observe first an increase in the frequency and then evidence of possible saturation. The decay in Newtonian fluids follows a power law, $\omega \sim 1/\sqrt{\mu}$ (dashed line). (b) The net swimming speed $U$ averaged over many oscillations is shown as a function of $\mu$. For both Newtonian and viscoelastic cases, the speed decays with increasing viscosity consistent with low Reynolds number motion where drag forces scale with viscosity and speed linearly. The error bars denote standard error from 10-20 sample individuals whose signed speed frequency spectrum has a clear single peak, evidence of active in-phase beating. Typically, more than 20 periods are used for each individual.}
\label{fig:2}
\end{figure*}
\begin{figure*}[t]
\centering\includegraphics[width=7.0cm]{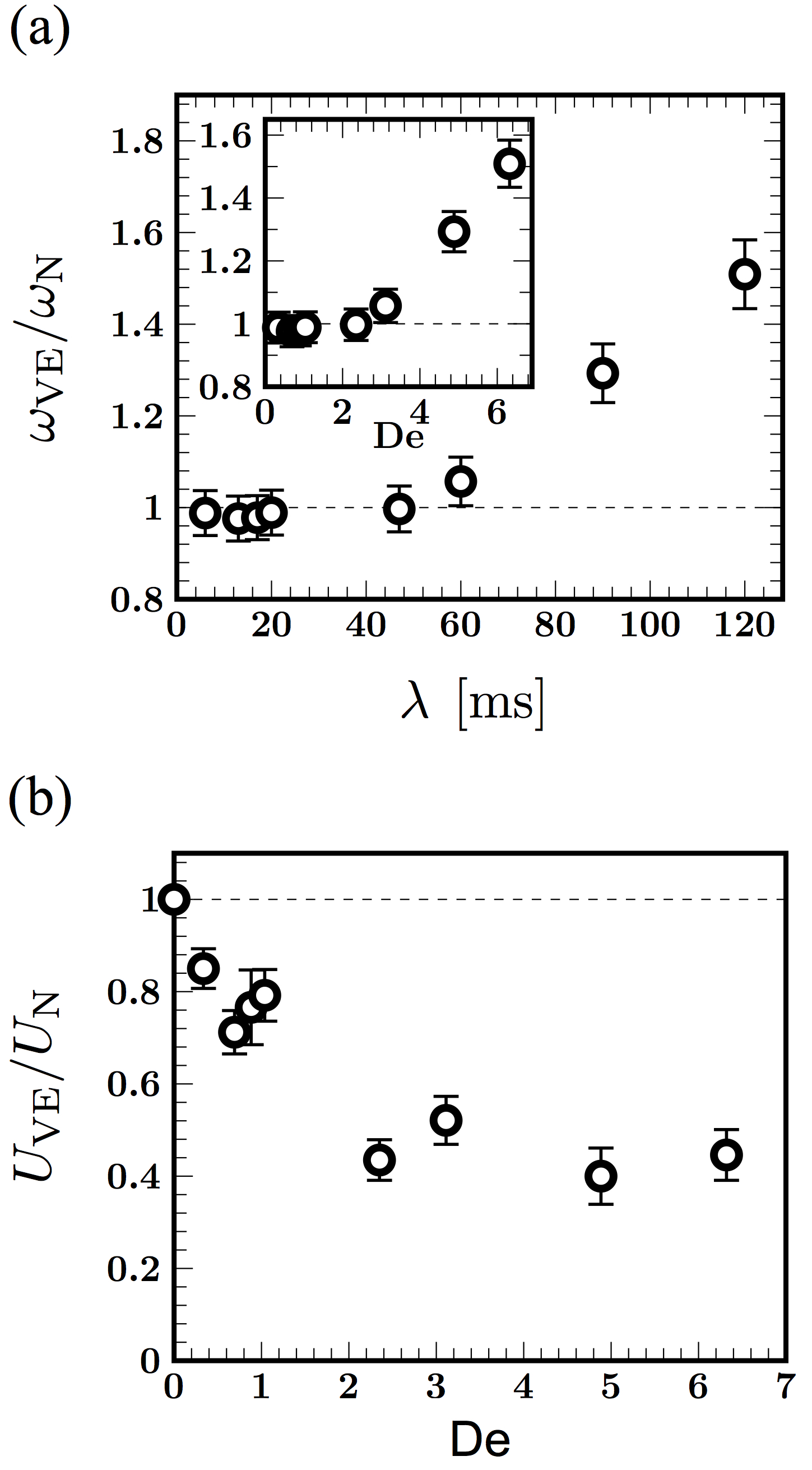}
\caption{
{\bf Fluid elasticity increases frequency but hinders motility.} { Experimental data replotted to emphasize the role of elasticity encapsulated by the Deborah number, De$\equiv \omega \lambda$, on the beating frequency and net swimming speed. (a) Frequency contrast $\omega_{\mathrm{VE}}/\omega_{\mathrm{N}}$ as a function of the the relaxation time $\lambda$ - error bars are representative of variations observed in the beating frequency in the viscoelastic fluid. The Newtonian value is based on the mean beating frequency in a fluid with the same viscosity. (Inset) The frequency contrast replotted as a function of the Deborah number. For De $\leq 2$ the ratio is close to unity. We then observe a nearly 50\% increase in the frequency as the Deborah number increases from 2 to 6.}  (b) The speed ratio $U_{\mathrm{VE}}/U_{\mathrm{N}}$ as a function of De - here, $U_{\mathrm{VE}}$ is the speed in viscoelastic fluids and $U_{N}$ is the swimming speed in a Newtonian counterpart.}
\label{fig:3}
\end{figure*}
\begin{figure*}[t]
\centering\includegraphics[width=\columnwidth]{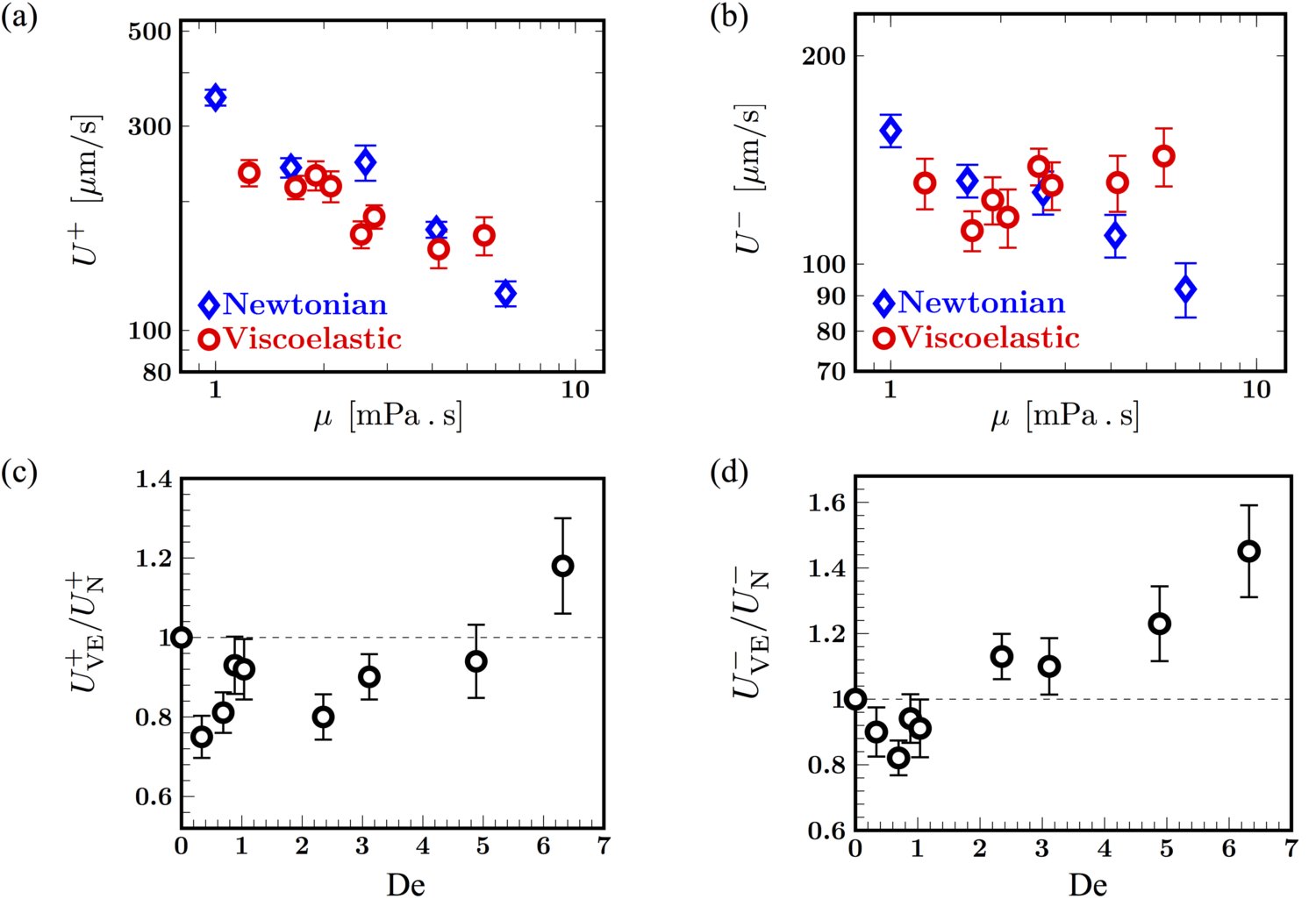}
\caption{
(Color Online) {\bf The effect of viscosity and elasticity on the swimming speed during power and recovery strokes analyzed separately.} Shown are data for  Newtonian (blue, diamonds) and viscoelastic fluids (red, circles). (a) The average speed $U^{+}$ during the execution of a power stroke. (b) The average speed $U^{-}$ during the recovery stroke. As before, the error bars denote standard error from 10-20 sample individuals. As viscosity increases, the Newtonian swimming speed decreases monotonically during both the power and recovery strokes. This contrasts significantly with two opposing trends seen in viscoelastic case. While  $U^+$ during the power stroke reduces with increasing viscosity similar to the Newtonian case, the speed during the recovery stroke, $U^-$,  is nearly constant and in fact increases for large viscosities. (c) Viscoelastic power stroke swimming speed normalized by the Newtonian counterpart of comparable viscosity, as a function of Deborah number. (d) The corresponding normalized speed for the recovery stroke.  
}
\label{fig:4}
\end{figure*}
\begin{figure*}[t]
\centering\includegraphics[width=\columnwidth]{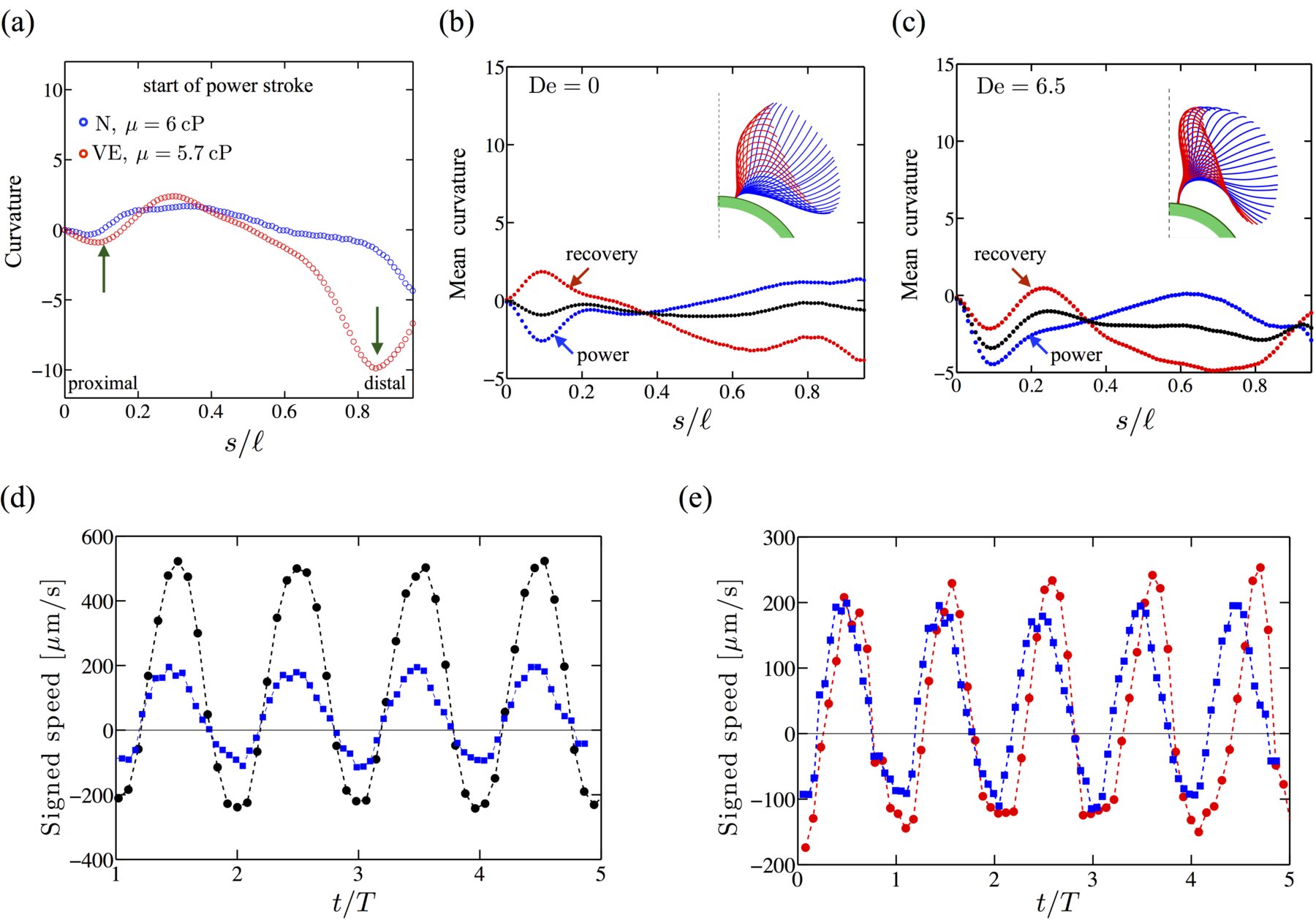}
\caption{ 
(Color Online) {\bf Structure of the flagellar stroke changes in viscoelastic fluids.} (a) Curvature of the flagella at the start of the power stroke in Newtonian ($\mu=6$mPa$\cdot$s ) and polymeric fluid (De = $6.5$ and $\mu=5.7$ mPa$\cdot$s ). Panels (b,c) Mean curvature profiles averaged over the power (P, blue) stroke, recovery (R, red)  stroke and full beat cycle (black) for the same fluids. (b) Newtonian fluid, and (c) Polymeric fluid. Panel (d) instantaneous speed as a function of time scaled with the period, $t/T$ for swimming in two Newtonian fluids -  moderate viscosity 2.6 mPa$\cdot$s (black, circles) and high viscosity 6 mPa$\cdot$s (blue, squares). (e) Instantaneous speed in a viscoelastic fluid corresponding to De $=6.5$ and $\mu=5.7 $mPa$\cdot$s (red, circles) and a Newtonian fluid of  viscosity $\mu=6 $mPa$\cdot$s (blue, squares). 
}
\label{fig:5}
\end{figure*}
\newpage
\end{document}